\documentclass[]{spie}  

\usepackage{amsmath,amsfonts,amssymb}
\usepackage{graphicx}
\usepackage[colorlinks=true, allcolors=blue]{hyperref}
\usepackage{makecell}
\usepackage{colortbl}
\usepackage{array}
\usepackage{afterpage}

\usepackage[table]{xcolor}

\def\arcdeg{\mbox{$^\circ$}}%
\def\elec{\mbox{e$^-$}}

\definecolor{eric}{rgb}{0.7, 0.7, 1.0}

\newcommand\nodata{ ~$\cdots$~ }%
\newcommand\nd{\nodata}

\title{Curving X-ray detectors for astrophysics applications}

\author[a]{Eric D.\ Miller}
\author[b]{James A.\ Gregory}
\author[b]{Keith Warner}
\author[a]{Beverly LaMarr}
\author[a]{Gregory Prigozhin}
\author[a]{Marshall W.\ Bautz}
\author[b]{Harry R.\ Clark}
\author[b]{Michael J.\ Cooper}
\author[b]{Kevan A.\ Donlon}
\author[a]{Catherine E.\ Grant}
\author[b]{WeiLin Hu}
\author[b]{Mallory A.\ Jensen}
\author[a]{Jill Juneau}
\author[b]{Renee D.\ Lambert}
\author[b]{Christopher W.\ Leitz}
\author[b]{David Volfson}
\author[b]{Douglas J.\ Young}

\affil[a]{Kavli Institute for Astrophysics and Space Research, Massachusetts Institute of Technology, Cambridge, MA, USA}
\affil[b]{Lincoln Laboratory, Massachusetts Institute of Technology, Lexington, MA, USA}

\authorinfo{Further author information:\\E.~D.~Miller: E-mail: milleric@mit.edu}

\newcommand{\fevv}{$^{\rm{55}}$Fe\ }
\newcommand{\procspie}{Proceedings of the SPIE}

\begin{document} 
\maketitle

\begin{abstract}
Next-generation X-ray optics will revolutionize high-energy astrophysics, yet they present several challenges to design a complementary focal plane. In particular, the focal surface is curved, requiring many small, flat sensors to achieve a large field. We present work building on MIT Lincoln Laboratory technology to curve the sensor itself, improving image quality and reducing complexity. Applying this technology to back-illuminated, large-format CCDs having well-characterized X-ray response, we describe the process and report success curving functional BI CCDs to a 2.5-m radius of curvature, achieving RMS curvature deviations less than 1 µm. We confirm that there is no appreciable increase in dark current and that the spectroscopic performance across the 0.3–6 keV band remains excellent. These results demonstrate that curved, large-format X-ray sensors are realizable, and the process can be extended to silicon detectors with other architectures, including active pixel sensors.
\end{abstract}


\section{INTRODUCTION}
\label{sec:intro}  

Future X-ray imaging missions require high spatial resolution over a wide field of view to enable deep surveys and reveal sources of multi-messenger signals without suffering from source confusion. Advanced X-ray optics are in development to provide these capabilities, along with high throughput across a broad X-ray band\cite{Zhangetal2019}. An inherent feature of grazing incidence X-ray optics is that the focal surface is curved, with radii as small as 1 to 2 m.  To fully exploit the capabilities of these optics and to maximize the science results, the camera must provide detectors that closely conform to curved focal surface. 

\begin{figure}[t]
\begin{center}
\includegraphics[width=.8\linewidth]{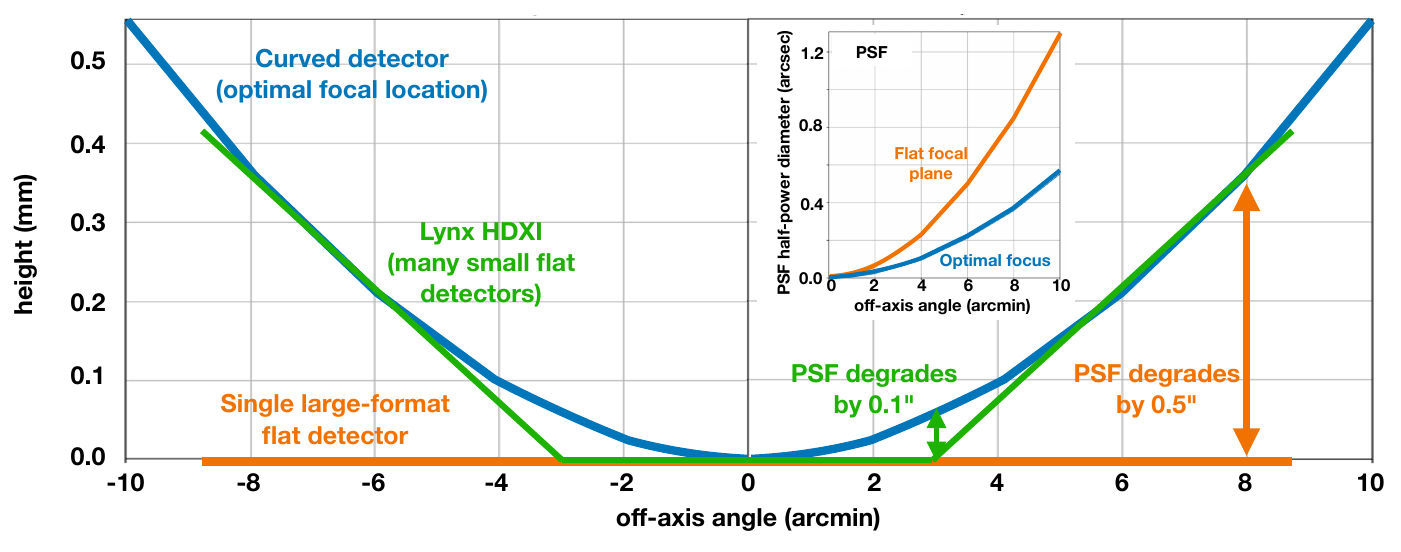}
\end{center}
\caption{Schematic of an advanced X-ray imager focal plane, placing three different detector layouts at the on-axis optimal focus. The inset compares the PSF of a flat focal plane to the optimal curved focal surface, assuming perfectly smooth and aligned mirror components and ignoring diffraction. Only a curved detector recovers the full resolving power of the optics at all off-axis angles.}
\label{fig:surface}
\end{figure}

To deal with this, most instrument designs employ multiple flat detectors tilted to follow the focal surface. Such was the case for the ACIS instrument flying on Chandra\cite{Garmireetal2003}, and the High-Speed Camera designed for the AXIS Probe mission concept\cite{Miller2023_AXIS,Miller2025_SPIE}. This is also the current plan for the High-Definition X-ray Imager (HDXI)\cite{HDXI} on the Lynx Flagship concept\cite{Lynx}. The effects of focal surface on image quality are shown in Figure \ref{fig:surface}, which results from detailed geometrical ray-tracing of a typical mirror shell of the Lynx mirror assembly. Flying a completely flat focal plane in an instrument like the HDXI on Lynx would double the PSF size over half of the FoV, hence the plan to tilt and tile detectors. However, Lynx’s need for a much larger area of sub-arcsecond imaging compared to Chandra, with similar focal length and much smaller depth of focus, requires a large number of small, four-side abutable detectors. This requirement complicates instrument design and introduces undesirable gaps in precious high-resolution field coverage. Use of a curved detector would greatly reduce this technical complexity and maximize the return of the expensive, advanced optics.

Here we provide updates from a joint effort between the MIT Kavli Institute for Astrophysics and Space Research (MKI) and MIT Lincoln Laboratory (MIT-LL) to demonstrate a technique for curving functional silicon X-ray detectors with no loss of spectral imaging performance. In previous work\cite{Miller2024_curvedccds}, we described the curving technique and demonstrated it on non-functional silicon dummies. That work also enabled refinement of the curving procedure and of the wafer thinning process used to fabricate back-illuminated (BI) charge-coupled devices (CCDs) with the support structure necessary to survive curving and mounting. Here we show successful curving of two functional, large-format BI CCDs, along with X-ray performance measurements. Compared with results from planar CCDs presented in our previous work, we conclude that the curving procedure has not affected the X-ray spectral performance of these CCDs in any measureable way.

\section{IMPLEMENTATION}

Curved detector technology has been demonstrated in the ground-based DARPA/US Air Force Space Surveillance Telescope (SST), a system with spherically curved MIT-LL detectors of 5.44-m convex radius of curvature operating in the visible spectrum. The curved sensor technology used for SST was made possible by development of thinned, BI CCDs\cite{Westhoff2009,Burke2007}. We and others have since demonstrated that an imager can be fabricated in the planar state and then deformed to a curved surface\cite{Gregory2015,Gaschet2019}, and curved CCD and CMOS sensors have also appeared commercially\cite{Itonaga2014,Guenter2017,Joaquina2022}. Challenges in fabricating curved sensors include mechanical deformation of the brittle silicon or other semiconductor detector material exceeding the fracture strain, introducing plastic deformation in metal layers, or buckling the deformed membrane into a shape that does not conform to the desired focal surface. Deformation of the silicon can raise the dark current of the device, constituting a further challenge\cite{Gaschet2019}. As we described previously\cite{Miller2024_curvedccds} and summarize below, the curving procedure requires mechanical contact with the CCD entrance window, which could damage the sensitive surface passivation.

\subsection{Project objectives}
\label{sect:objectives}

The objectives of this project are to:

\begin{itemize}
    \item Curve an existing back-illuminated (BI) X-ray CCD (the MIT-LL CCID94), with diagonal length of 65 mm and active (fully depleted) thickness of 100 $\mu$m, to a radius of curvature of 2.5 m with an accuracy of 5 $\mu$m;
    \item Characterize detector properties (cosmetics, dark current, noise, responsivity, charge-transfer efficiency, and X-ray spectral resolution) of both planar and curved CCID94s to identify performance changes resulting from the deformation process; and  
    \item Expose planar and curved CCID94 detectors to notional vibration levels and representative thermal and radiation environments.  
\end{itemize}	

\begin{figure}[t]
\begin{center}
\includegraphics[height=2.3in]{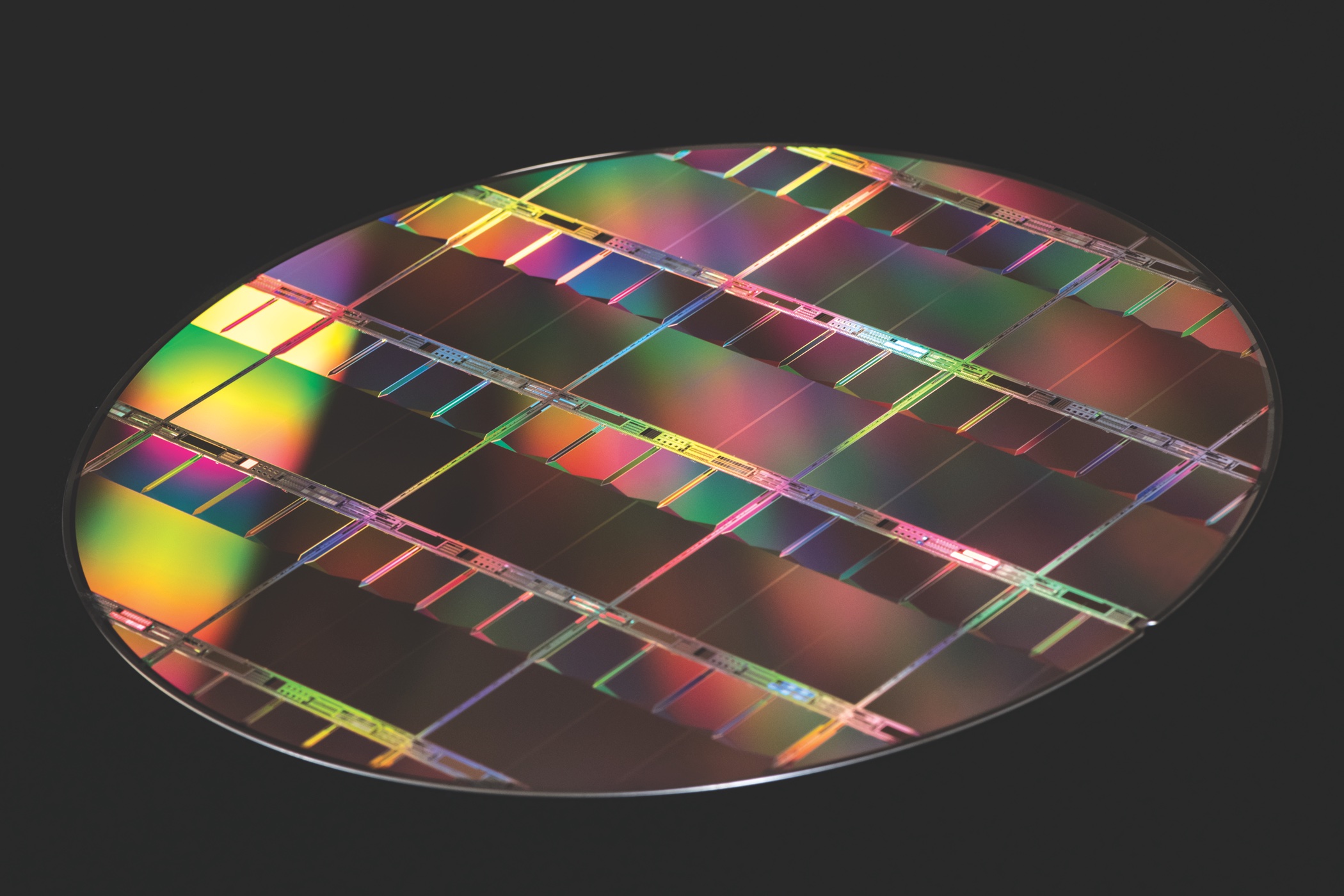}
\includegraphics[height=2.3in]{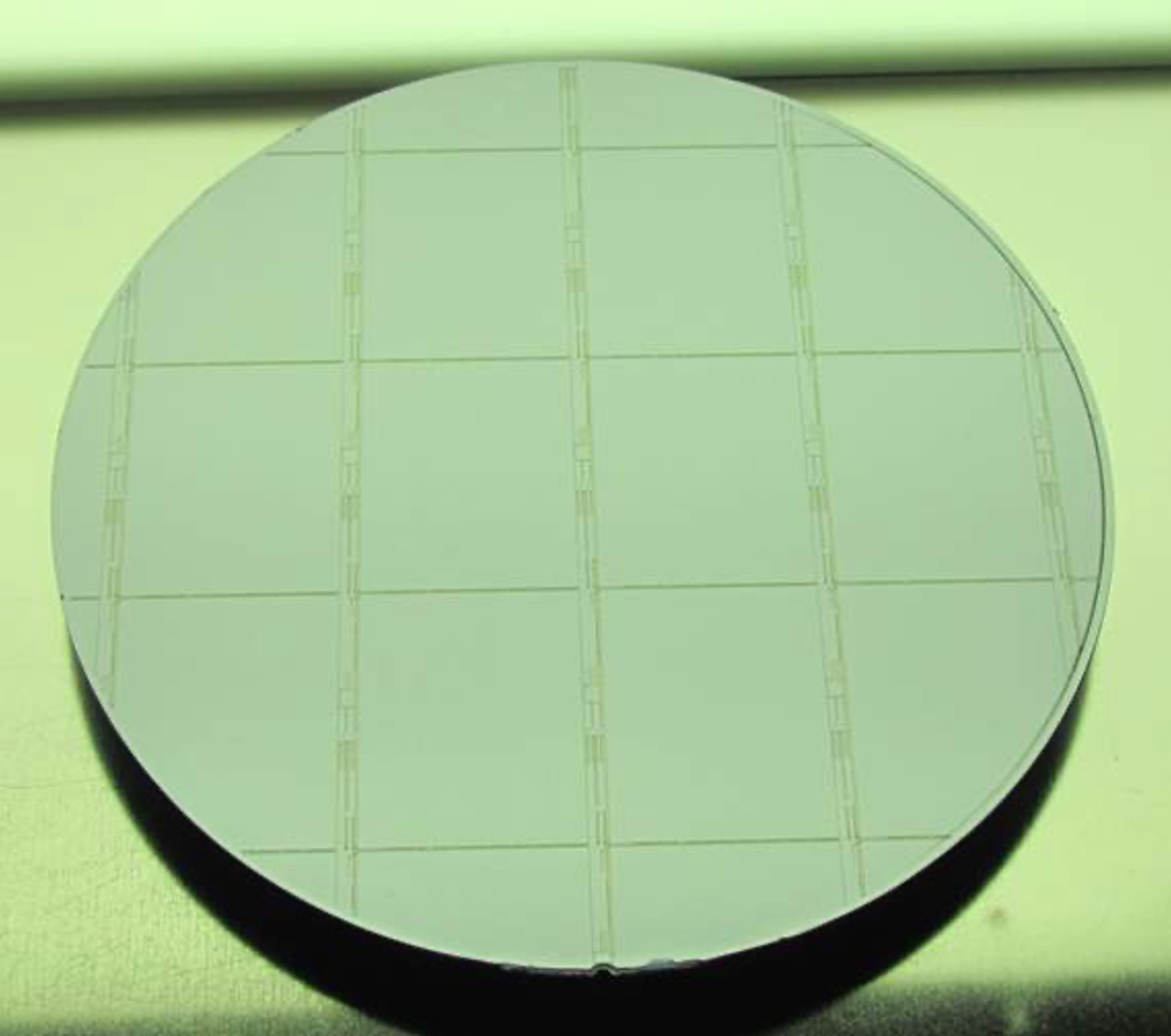}
\end{center}
\caption{Photographs of 200-mm wafers containing CCID94 CCDs through front-illumination (left) and back-illumination (right). Eight complete CCDs can be seen, each with a 50$\times$25-mm rectangular imaging area and four distinct framestore regions; these differing regions in the CCD can only be seen  in the front-illuminated devices. Each framestore feeds two outputs.}
\label{fig:ccid94_photo}
\end{figure}

For a demonstration target and to choose a test detector, we adopt notional requirements of the Lynx HDXI design\cite{Lynx,HDXI}, a detector array about 65-mm across with focal-surface radius of curvature $\sim$2.5 m. The CCID94 detector developed by MIT-LL provides 2048 columns and 1024 rows of 24-$\mu$m pixels in the imaging area, or 49.1$\times$24.6 mm. The entire die including the adjacent framestore (with equal pixel count but smaller pixels) is 50.7$\times$40.5 mm, or $\sim$65 mm along the diagonal. This diagonal extent is nearly identical to that for a square 33$\times$33-mm imaging-area chip, and so it provides a useful prototype on which to test the curving procedure. The CCID94, summarized in Table \ref{tab:ccid94}, has undergone years of development and fabrication at MIT-LL and extensive X-ray testing at MKI, and it is a well-understood, high-performance imager. It was baselined for the focal plane of the high-resolution X-ray grating spectrometer on the Arcus Explorer and Probe class mission concepts\cite{Smith2020_Arcus,Smith2023_Arcus,Smith2024_Arcus,Grant2024_Arcus}. Photographs of 200-mm wafers of CCID94 devices are shown in Figure \ref{fig:ccid94_photo}.

\begin{table}[t]
\caption{Features of the MIT-LL CCID94 CCD}\label{tab:ccid94}
\begin{center}       
\small
\begin{tabular}{|l|c|}
\hline\hline
\textbf{Feature}          & \textbf{CCID94} \\ \hline
Format                    & Frame-transfer, 2048$\times$1024 pixel imaging array \\
Image area pixel size     & 24$\times$24 $\mu$m \\
Output ports              & 8 pJFET                   \\
Transfer gate design      & Triple layer polysilicon  \\
Additional features       & Trough, charge injection  \\
BI detector thickness     & 50--100 $\mu$m            \\
Back surface              & MIT-LL MBE$^a$ 5--10 nm          \\
Typical serial rate       & 0.5 MHz                   \\
Typical parallel rate     & 0.1 MHz                   \\
Full-frame read time      & $<$1 s                       \\
\hline\hline
\multicolumn{2}{l}{$^a$Molecular beam epitaxy, see Section \ref{sect:curving_procedure}.} \\
\end{tabular}
\end{center}
\end{table} 

\subsection{Curving procedure}
\label{sect:curving_procedure}

\begin{figure}[t]
\begin{center}
\vspace*{\baselineskip}
\includegraphics[width=\linewidth]{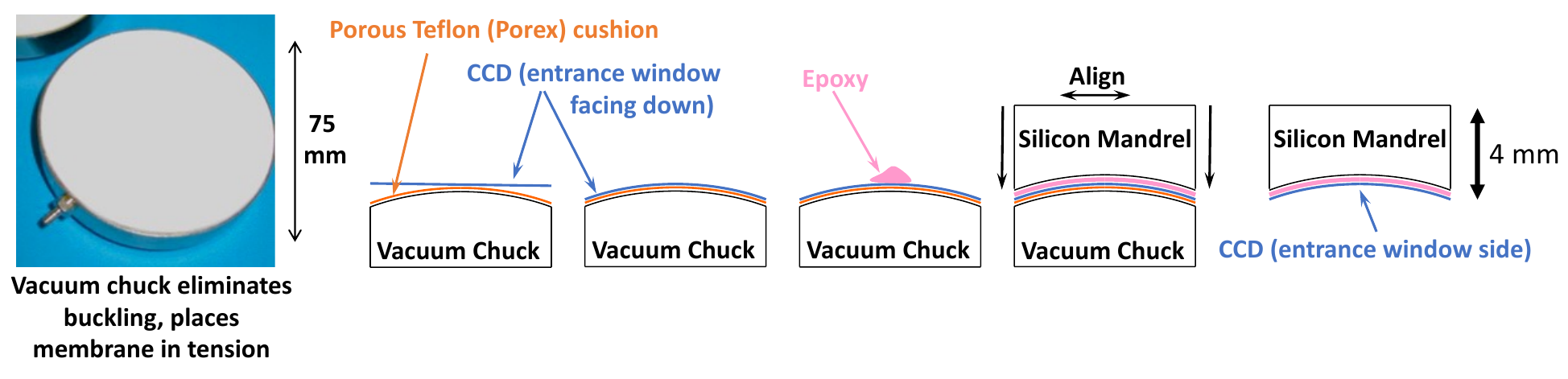}
\end{center}
\caption{Schematic of the curving process for a diced CCD. The steps are described in the text.}
\label{fig:curving}
\end{figure}

The technique to prepare CCDs for curving is detailed in our previous work\cite{Miller2024_curvedccds}. Here we describe the curving procedure we used, which has some alterations from the plan we described previously---that plan was adapted from the SST method to apply to a concave curvature, and it is reproduced in Figure \ref{fig:curving}. First a protective porous Teflon foil (``Porex'', shown as an orange layer) is placed on the convex, milled portion of a ceramic vacuum chuck, which is convex with the desired radius of curvature. The foil protects the illuminated surface of the CCD. The CCD is placed illuminated-side down over the chuck and a vacuum applied, drawing the CCD into the appropriate curvature. Epoxy is placed on the now-convex rear surface of the thinned handle silicon and a  concave silicon mandrel is brought into contact and aligned using a specialized jig. The mandrel and silicon are brought together with a  modest and non-destructive force, using a hydraulic plunger, and maintained for several days to ensure proper curing of the epoxy between the imager and the mandrel. The resultant curved CCD is permanently mounted on its mandrel.

\begin{figure}[p]
\begin{center}
\vspace*{\baselineskip}
\includegraphics[width=.60\linewidth]{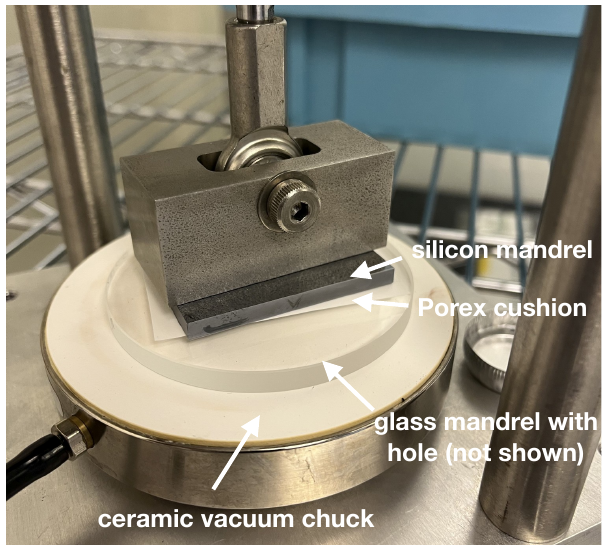}
\end{center}
\caption{Photograph of the updated curving apparatus with components identified.}
\label{fig:curving_photo}
\end{figure}

As reported previously, the vacuum chuck for this project was fabricated with the wrong radius of curvature, 2250 mm. The reason for this error is under investigation, and we were unable to fabricate a replacement chuck with the correct figure. As a result, we devised a modified curving method. Briefly, a hole was drilled in one of the glass mandrels that were fabricated with the correct radius of curvature for testing purposes, as described previously\cite{Miller2024_curvedccds}. The thinned, diced BI CCD was place on the convex side of this mandrel atop the Porex cushion, and vacuum was applied to the other side of the mandrel hole using the incorrectly figured chuck. This successfully conformed the CCD to the mandrel shape. Then, as originally planned, the silicon support mandrel was epoxied to the convex side of the curved CCD and allowed to cure. A photograph of the curving setup is shown in Figure \ref{fig:curving_photo}. 

Three CCID94 BI devices were curved in this manner; two (identified from the wafer and chip location as W201C7 and W202C1) were selected pre-dicing as flight-quality devices, projecting good X-ray performance. The third (W202C7) had qualities of an engineering development unit (EDU), with an expectation of basic functioning but likely poor performance. The two wafers these devices came from had been mounted to a standard Si support wafer with epoxy, the CCD wafer was thinned to 100 µm, and the support wafer was then thinned to approximately 150 µm, resulting in a Si sandwich approximately 250 µm thick. For backside passivation\cite{Miller2024_curvedccds}, wafer 201 received a 5-nm molecular beam epitaxy (MBE) layer, while wafer 202 received a thicker 10-nm layer.  The MBE layer\cite{Ryuetal2018} is a silicon film, uniformly-doped with boron to a doping level of about $4\times10^{20}$ cm$^{-3}$.  Both wafers received a 2.5-nm atomic layer deposition (ALD) of aluminum oxide to protect the thin, highly doped MBE layer. A summary of the processing applied to each detector is shown in Table \ref{tab:ccds}, including the planar device we discussed previously\cite{Miller2024_curvedccds}.

\begin{table}[p]
\caption{MIT-LL CCDs used in this study.}\label{tab:ccds}
\begin{center}       
\small
\begin{tabular}{|l|c|c|c|c|}
\hline\hline
\textbf{CCD device} & \textbf{W19C5} & \textbf{W202C1} & \textbf{W201C7} & \textbf{W202C7}\\ 
MKI designation & CCID94L1W19C5 & CCID94W202C1 & CCID94W201C7 & CCID94W202C7 \\ 
Curvature       &  planar       &  curved       &  curved       &  curved     \\ 
Depletion depth & 50 µm         & 100 µm        & 100 µm        & 100 µm \\ 
MBE treatment   & 10 nm         &  10 nm        &   5 nm        &  10 nm  \\
Wafer-level quality   & flight         &  flight        &   flight        &  EDU  \\
\hline
\hline
\end{tabular}
\end{center}
\end{table} 

\begin{figure}[p]
\begin{center}
\vspace*{\baselineskip}
\includegraphics[width=\linewidth]{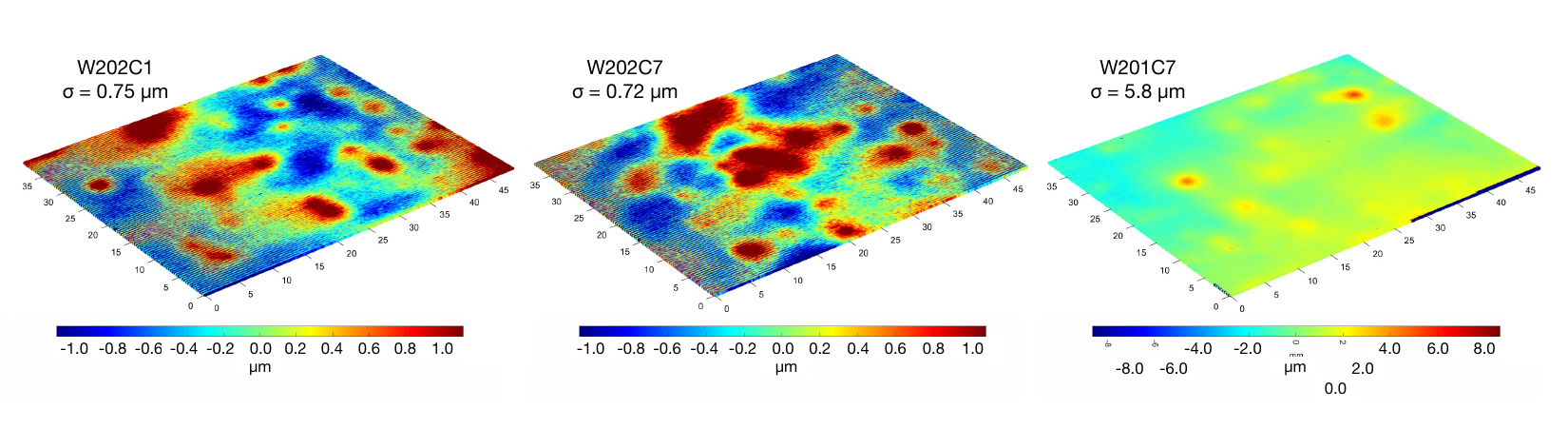}
\end{center}
\caption{Residual profilometry measurements of each curved sensor after subtracting a sphere of 2.5-m radius. The two sensors on the left easily meet the required RMS deviations of 5 µm. The surface deviations of the right-most sensor, W201C7, are dominated by a small area near the edge of the chip.}
\label{fig:profiles}
\end{figure} 

Profilometry measurements performed at MIT-LL confirm that the CCDs as mounted in their mandrels meet the design radius of curvature of 2.5 m. Figure \ref{fig:profiles} shows surface residuals after subtracting a sphere of that radius from the measurement for each device. W202C1 and W202C7 easily meet the RMS surface deviation requirement of 5 µm. W201C7 has a larger surface deviation of $\sigma = 5.8$ µm, dominated by a small region near the edge that may result from a measurement error or a non-uniformity in the thinned 100-µm layer. Most of the chip has deviations much smaller than this. In addition, our choice of surface roughness requirement is somewhat arbitrary, selected to be significantly smaller than both the depletion depth of the devices (100 µm) and the magnitude of deviations from the focal surface expected by tiling the HDXI surface with small, planar devices (also $\sim$100 µm, see Figure \ref{fig:surface}).

\begin{figure}[t]
\begin{center}
\vspace*{\baselineskip}
\includegraphics[height=.40\linewidth]{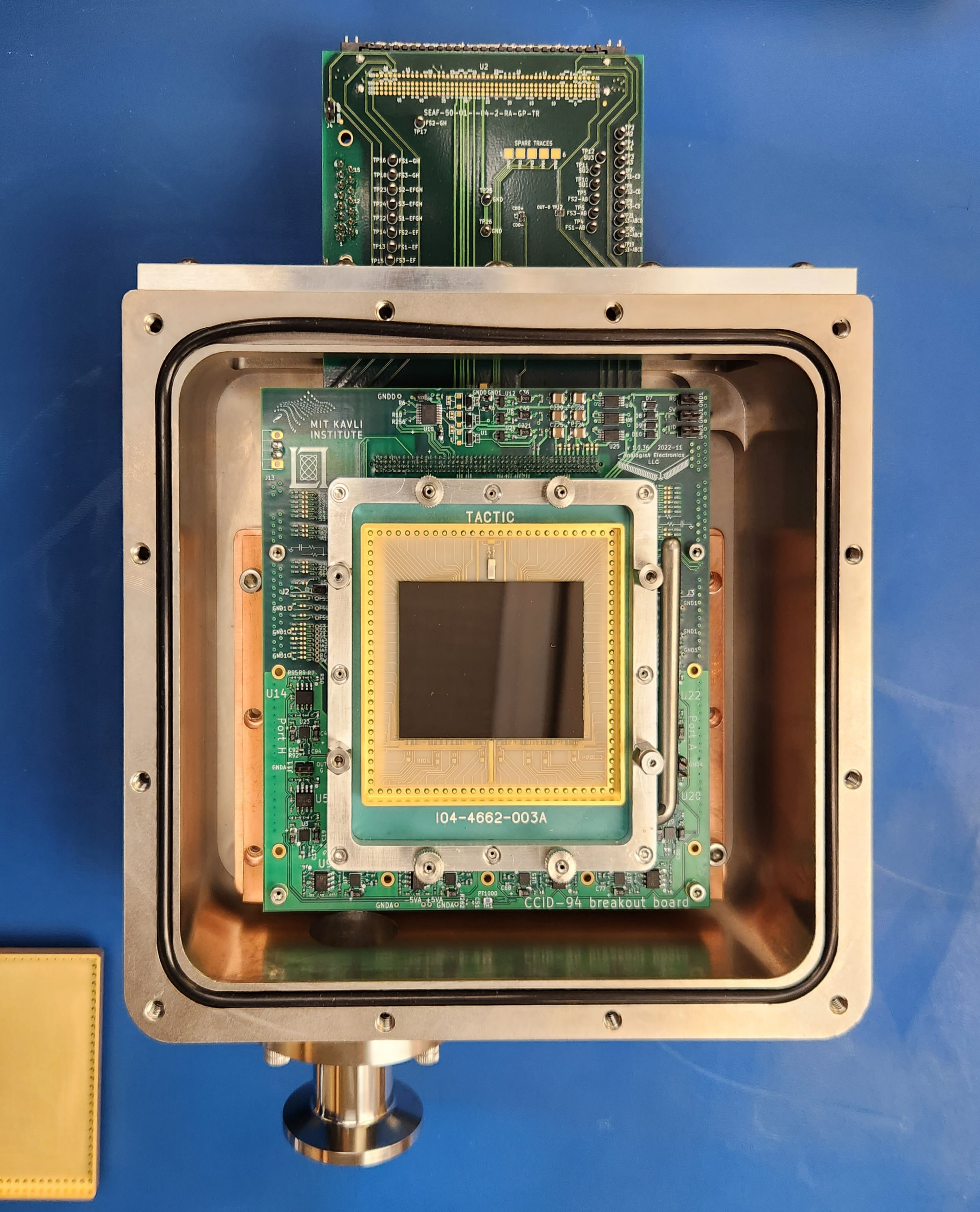}
\includegraphics[height=.40\linewidth]{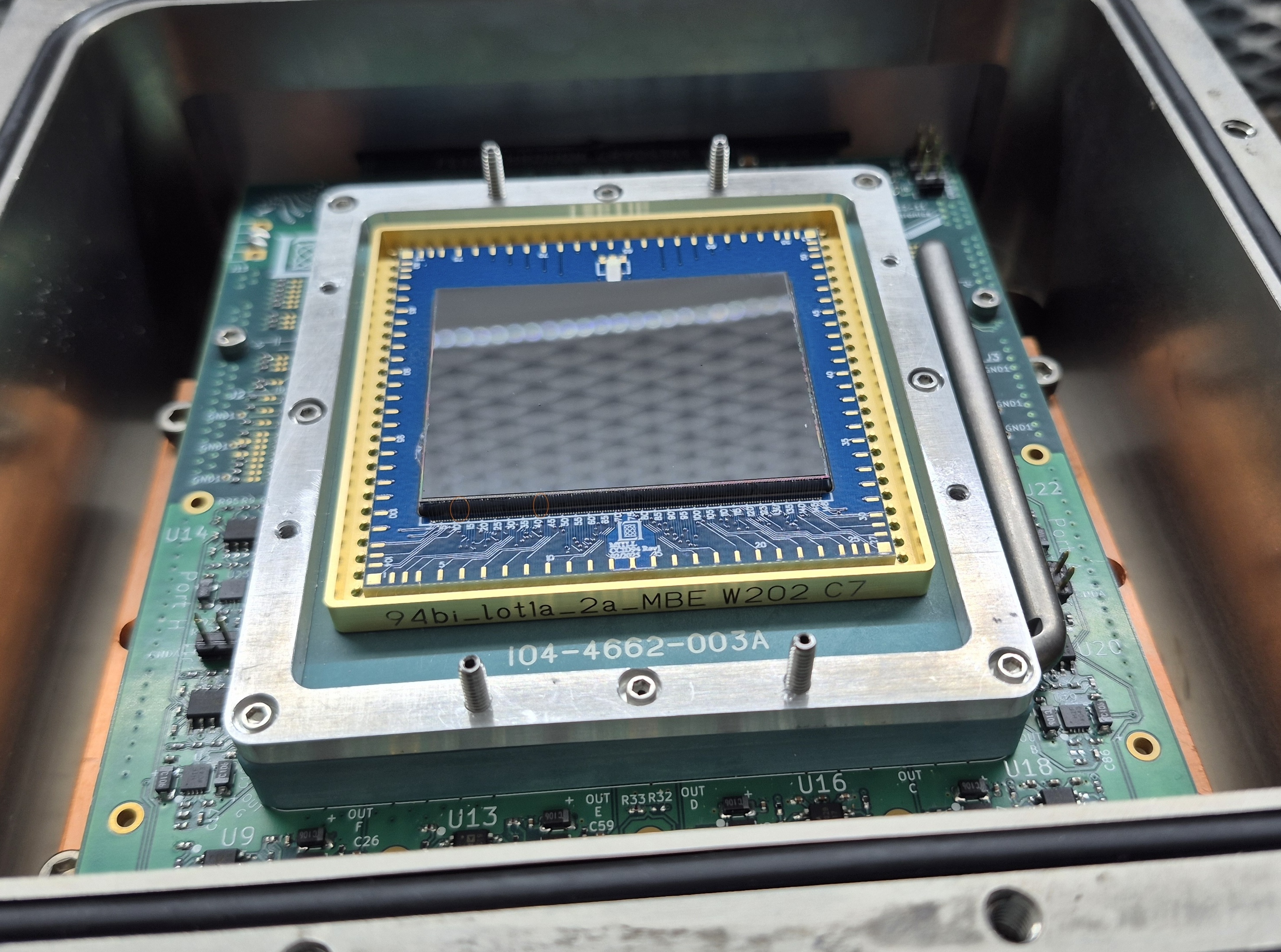}
\end{center}
\caption{Photographs of a (left) planar and (right) curved BI CCID94 CCD detector in the vacuum housing in preparation for X-ray performance testing. The slight concave curvature of the device in the right panel can be seen in the reflection of the white fluorescent light, which would otherwise be straight, and in distortion of the reflected rectangular grid.}
\label{fig:detectors_in_housing}
\end{figure}

Each of the three curved devices was installed with an interposer in a 104-pin Kovar package and delivered to MKI's X-ray Detector Lab facilities on MIT's main campus for X-ray performance testing. Figure \ref{fig:detectors_in_housing} shows photographs of a planar and curved CCD in the vacuum housing used for testing. The surface of the curved CCD is clearly not planar, as the reflection of an overhead fluorescent shows a curved image.

\section{X-ray performance testing}
\label{sect:testing}

The X-ray performance testing goals, the test facilities, and the data processing and analysis techniques are described in great detail in our previous work\cite{Miller2024_curvedccds}, which also presents comparison results from the planar, 50-µm-thick BI CCID94 designated W19C5. The performance properties we aim to measure are those that are most susceptible to damage caused in the curving process, namely readout noise (measured at a number of serial rates), dark current (measured as a function of temperature), and spectral performance (quantified as the FWHM response to monochromatic X-ray emission). The response measurement is especially important in the soft energy band, which is sensitive to damage to the backside entrance window passivation. For these purposes, we utilize a radioactive $^{55}$Fe source for full-frame illumination with Mn K$\alpha$ (5.9 keV) and K$\beta$ (6.4 keV) X-rays; and an In-Focus Monochromator (IFM)\cite{Hettrick1990_ifm} that uses grazing incidence reflection gratings to produce clean monochromatic lines at energies below 2 keV. The typical spectral resolving power of the IFM is $\lambda/\Delta\lambda = E/\Delta E \sim $ 60--80, far higher than that of the CCD itself. The basic data product of a test run is an event list, which for every detected X-ray records the time, location, and pulse heights in a 3$\times$3 pixel island around each local maximum. Noise thresholds and event grades or pixel patterns are applied as previously described\cite{Miller2024_curvedccds}, using methods similar to those employed on the Chandra ACIS and Suzaku XIS instruments, which use similar heritage CCDs. In particular, in this work each event is assigned a pixel ``multiplicity'' n\# which encodes the number of pixels in the 3$\times$3 island that are above the neighbor noise (or ``split'') threshold. 

\subsection{Noise and gain}
\label{sect:noise}

Basic performance of both curved CCDs is good, and comparable to the planar device. This can be seen in Figure \ref{fig:94noise_gain}, which shows readout noise and gain at a readout speed of 0.5 MHz, corresponding to a 0.5-s frame time. All three detectors exhibit low noise that is uniform across outputs at the lowest temperatures and varies smoothly with temperature. The gain is also uniform and well behaved with temperature below about $-40$\arcdeg C, where dark current becomes low. There are some important exceptions and caveats to these results:

\begin{figure}[p]
\begin{center}
\includegraphics[width=.9\linewidth]{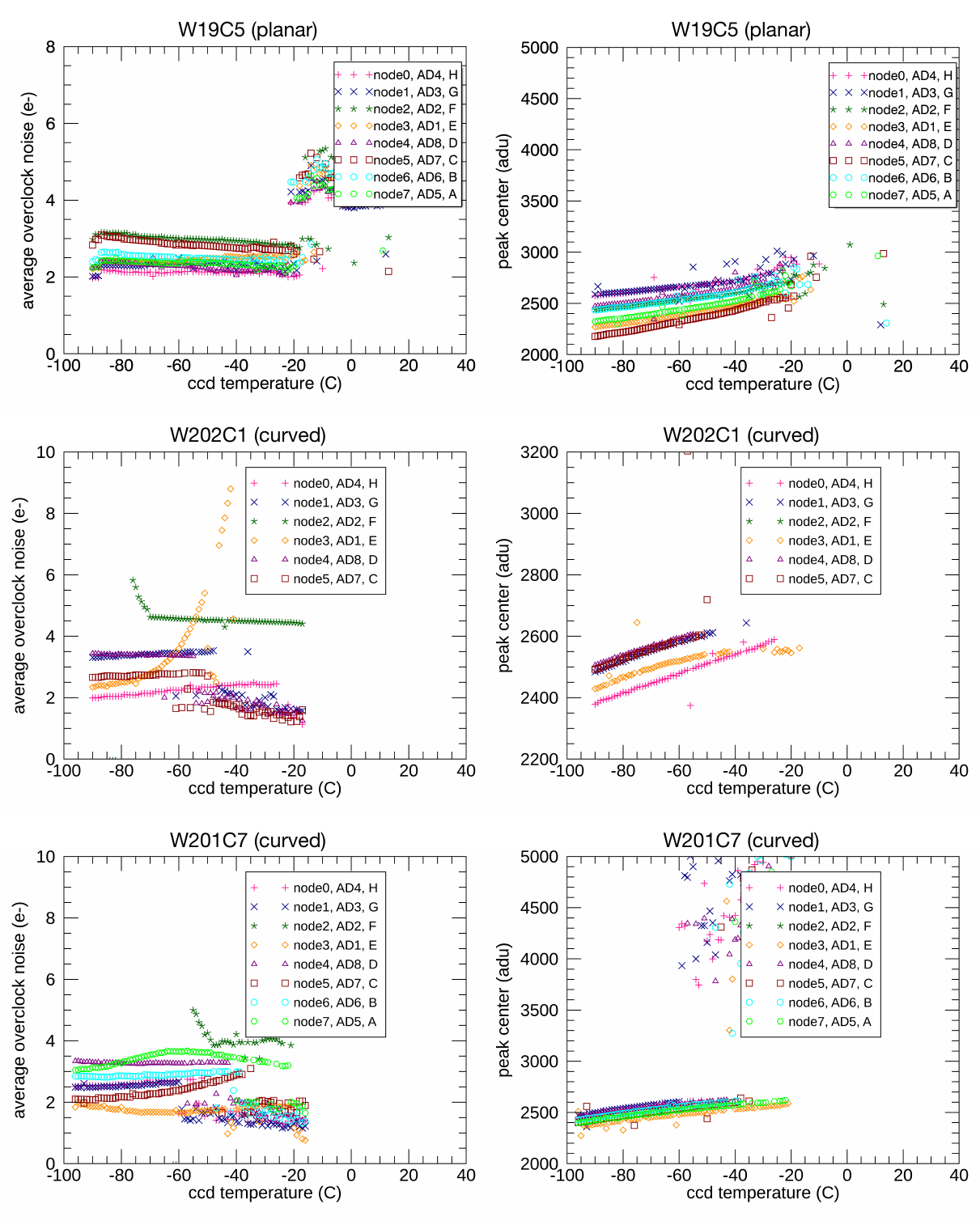}
\end{center}
\caption{(left) Readout noise as a function of temperature for all working output nodes on the CCDs. (right) Peak in instrument units (ADU) of the Mn K$\alpha$ line at 5.9 keV. Results are shown for (top) the planar W19C5, (middle) the curved W202C1, and (bottom) the curved W201C7.}
\label{fig:94noise_gain}
\end{figure} 

\begin{itemize}
    \item  On the first curved device, two outputs (A and B) stopped working shortly after we began testing. We were able to take some data with them through several thermal cycles, and their performance was in line with the other outputs, although the bias levels had not been tuned at that point. We believe their later failure is due to an unplanned low temperature excursion of the readout board rather than damage to the chip from curving. We are continuing to diagnose this issue.
    \item Some nodes on the curved devices exhibit different behavior with temperature compared to the flat device. In particular, output F has very high noise which is off the plot at low temperatures, and it has low gain which is off the plot at all temperatures. This could arise from the readout chain downstream of the CCD, perhaps in the board that is used for all testing, although the planar CCD did not show such behavior on this node. We are looking into this anomaly. There is generally more dispersion among nodes in the noise trends of the curved devices, although at expected operating temperatures $\leq-70\arcdeg$C, the readout noise of behaving nodes is below 4 \elec\ RMS in all devices.
    \item The gains derived from the curved CCDs show significant scatter to a much lower temperature ($\sim-50\arcdeg$C) than the planar device ($\sim-20\arcdeg$C). We suspect this scatter is due to dark current signal diluting the spectrum and affecting the automated fits to the Mn K lines produced by the \fevv source. This does not indicate more dark current in the curved devices, as we discuss below. The \fevv source used for the planar device was bright and mounted within the housing at a distance of a few cm, while the \fevv source illuminating the curved devices is fainter and mounted in the IFM at a distance of $\sim$1 m. The higher flux of Mn K photons on the planar device enables successful Gaussian fitting in the presence of more dark current, hence the more reliable results at higher temperatures.
\end{itemize}

\subsection{Dark current}
\label{sect:dark_current}

\begin{figure}[b]
\begin{center}
\includegraphics[width=\linewidth]{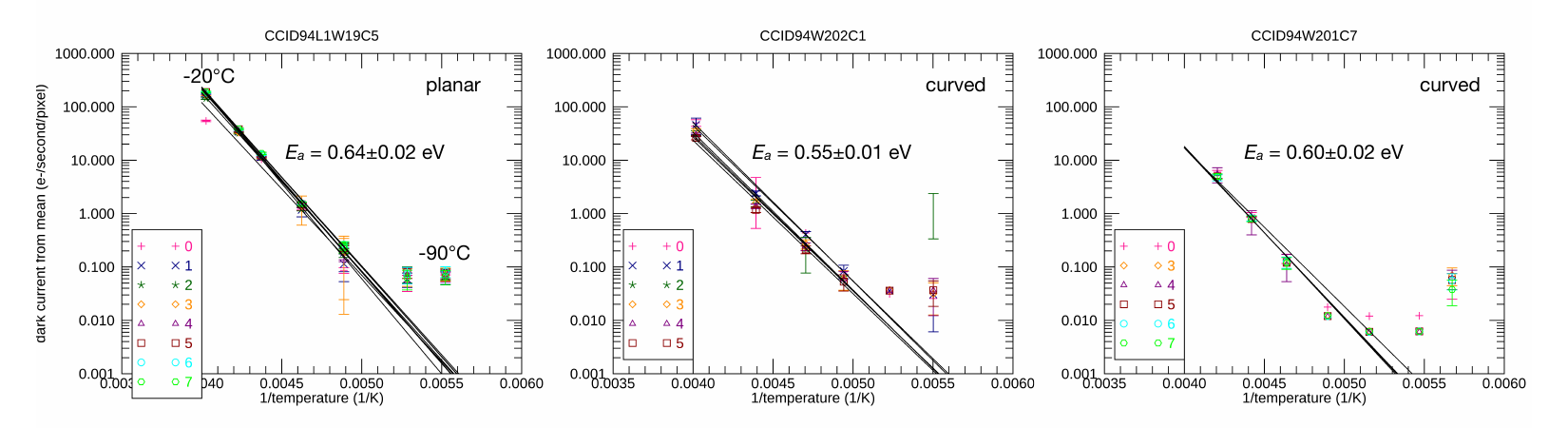}
\end{center}
\caption{Dark current as a function of temperature for all three CCID94s, with different output nodes in different colors. Including only temperatures where dark current is reliably measured, the temperature dependence follows an Arrhenius with activation energy as indicated. The lowest temperature divergent points are explained in the text.}
\label{fig:94dark_current}
\end{figure}

We measured the dark current as a function of temperature for each detector by taking a series of datasets with varying integration time and no illumination of the focal plane. For each integration time, the bias-corrected mean pixel level is calculated from a central region of each node, and then dark current is calculated from a linear fit to the mean pixel level vs.\ integration time at that temperature. The conversion from ADU to \elec\ uses the gain derived from fits to the Mn K$\alpha$ line at 5.9 keV (see Figure \ref{fig:94noise_gain}, right panels). The dark current trends are shown in Figure \ref{fig:94dark_current}.

At temperatures above about $-60\arcdeg$C, the dark current is well-fit by an Arrhenius law, $I_d \propto \exp(-E_a/kT)$, with activation energy $E_a \approx 0.6$ eV for all operating segments in all three detectors. This is close to the value of one-half the band-gap for silicon, as expected for thermal activation of electrons due to lattice defects; for MIT-LL CCDs, the activation energy for FI and BI devices generally falls in the range of 0.60 to 0.63 eV. The curved devices have a significantly lower activation energy than the planar one, however it is unclear whether this is a result of the curving process, wafer-to-wafer differences, or our measurement and analysis. Interestingly, the dark current is lower in the curved devices, especially at the highest temperatures. While the details of these differences require further exploration, it is clear that curving has not increased the dark current at the typical operating temperature of $\leq-70\arcdeg$C in a way that affects the performance of the device for X-ray spectral imaging.

The measured dark current in all three devices levels off at the coldest temperatures. We attribute this to a small amount of light leak in our test setup; the longest exposures have some very low level of positive signal, and as there is no apparent temperature dependence, this appears to arise from an unvarying source. The level that we see, $\leq0.1$ \elec\ s$^{-1}$ pix$^{-1}$, is equivalent to $\leq0.2$ \elec\ event$^{-1}$ for our 0.5-s frame time, well below the typical requirement for on-orbit light-blocking power of a flight instrument. For a ground testbed that lacks an optical blocking filter and has not been specifically designed for light-tightness in an active lab setting, this is actually quite good.

\subsection{Spectral response}
\label{sect:response}

\begin{table}[b]
\caption{CCD performance testing results.}\label{tab:results}
\begin{center}       
\small
\begin{tabular}{|p{.6in}p{.6in}|>{\centering}p{1.2in}|>{\centering}p{1.2in}|>{\centering\arraybackslash}p{1.2in}|}
\hline\hline
\multicolumn{2}{|l|}{\textbf{CCD device}} & \textbf{W19C5} & \textbf{W202C1} & \textbf{W201C7} \\ 
\multicolumn{2}{|l|}{Curvature}       &  planar     &  curved     &  curved     \\ 
\multicolumn{2}{|l|}{Representative node}     &  C     &  C     &  C     \\ 
\hline
\multicolumn{2}{|l|}{Detector temperature}    &  $-89\arcdeg$C    &  $-91\arcdeg$C    &  $-90\arcdeg$C    \\ 
\multicolumn{2}{|l|}{Serial readout rate}     &  0.5 MHz     &  0.5 MHz     &  0.5 MHz     \\ 
\multicolumn{2}{|l|}{Readout noise (RMS)}     &  2.1--3.1 \elec\ &  2.3--3.5 \elec\ &  1.8--3.3 \elec\ \\ 
\multicolumn{2}{|l|}{Parallel CTI$^a$}        &  $<10^{-6}$ & $<10^{-6}$ & $<10^{-6}$ \\ 
\hline
\multicolumn{5}{|l|}{Spectral resolution FWHM in eV (fraction of events)$^b$} \\ 
\hline
C K           & all         &  68 (100\%) &  68 (100\%) &  57 (100\%) \\
0.27 keV      & n1          &  65 (51\%)  &  58 (24\%)  &  48 (22\%)  \\
              & n2          &  70 (43\%)  &  64 (65\%)  &  51 (63\%)  \\
              & n3          &  75 (5\%)   &  58 (8\%)   &  52 (14\%)  \\
              & n4          &  60 (1\%)   &  39 (1\%)   &  46 (1\%)   \\
\hline                                                                           
O K           & all         &  57 (100\%) &  89 (100\%) &  71 (100\%) \\
0.53 keV      & n1          &  53 (39\%)  &  69 (8\%)   &  56 (6\%)   \\
              & n2          &  58 (48\%)  &  81 (47\%)  &  59 (45\%)  \\
              & n3          &  69 (8\%)   &  91 (28\%)  &  63 (28\%)  \\
              & n4          &  60 (5\%)   &  96 (16\%)  &  61 (19\%)  \\
\hline                                                                
Mn K          & all         & 139 (100\%) & 179 (100\%) & 134 (100\%) \\
5.9 keV       & n1          & 129 (18\%)  & 130 (2\%)   & 137 (4\%)   \\
              & n2          & 136 (44\%)  & 161 (12\%)  & 129 (11\%)  \\
              & n3          & 140 (18\%)  & 154 (16\%)  & 136 (16\%)  \\
              & n4          & 143 (20\%)  & 182 (48\%)  & 131 (57\%)  \\
              & n5          & \nd         & 199 (12\%)  & 145 (9\%)   \\
              & n6          & \nd         & 194 (11\%)  & 136 (2\%)   \\
\hline
\hline
\multicolumn{5}{l}{$^a$CTI is presented as fractional charge loss per transfer.} \\
\multicolumn{5}{l}{$^b$Spectral resolution is provided for each pixel multiplicity `n\#', with the fraction of events } \\
\multicolumn{5}{l}{~with that multiplicity given in parentheses. Multiplicities representing fewer than 1\% of } \\
\multicolumn{5}{l}{~events are not shown.} \\
\end{tabular}
\end{center}
\end{table} 

\begin{figure}[p]
\begin{center}
\includegraphics[width=.9\linewidth]{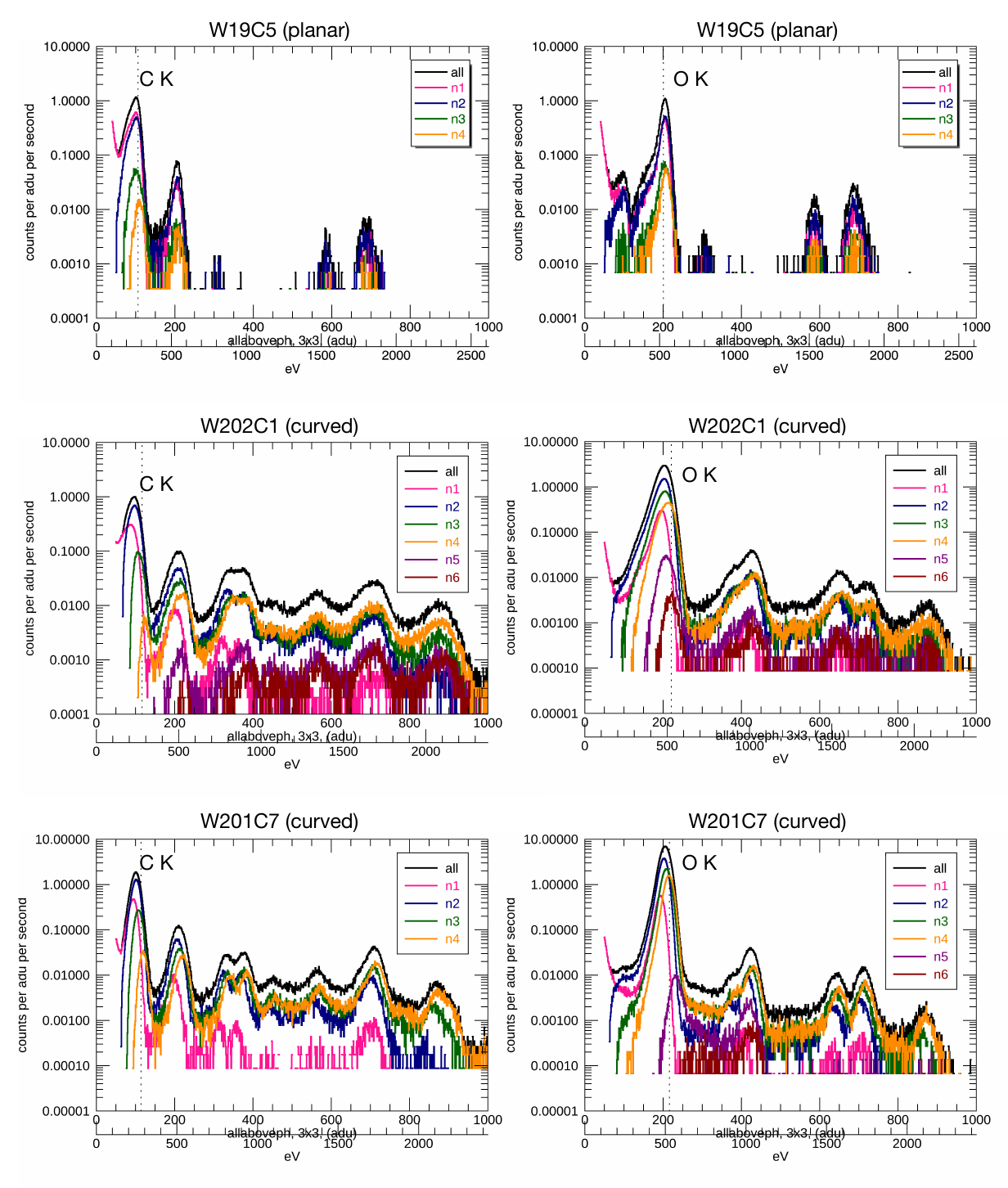}
\end{center}
\caption{Spectra of (left) C K and (right) O K for a representative output node on each of the three CCDs: (top) the planar W19C5, (middle) the curved W202C1, and (bottom) the curved W201C7. Data for the planar device was  obtained at the MKI polarimetry beamline\cite{Marshall2024_SPIE}, and additional lines from different grating orders can be seen at low energies, along with fluorescence lines of Al K$\alpha$ (1.49 keV) and W M$\alpha$ (1.78 keV) produced by the polarizing multilayer holder and the multilayer itself, respectively. The curved devices were measured in the IFM, and the spectra show lines from higher orders. The dashed vertical lines show the expected line centers. Different colored lines show contributions from events with different pixel multiplicity, as indicated in the legends and discussed in the text.}
\label{fig:94spectra}
\end{figure} 

The spectral response of all three devices was measured at three energies, 0.27 keV (C K), 0.53 keV (O K), and 5.9 keV (Mn K$\alpha$), at a temperature of about $-90\arcdeg$C. The results for a typical readout node are shown in Table \ref{tab:results}, and spectra for the C and O K measurements for the same nodes are shown in Figure \ref{fig:94spectra}. The results for the planar W19C7 device are taken from our previous work\cite{Miller2024_curvedccds}, which used the MKI polarimetry lab beamline\cite{Marshall2024_SPIE} for C and O K. The curved devices were measured in the IFM.

The performance of the curved devices is similar to that of the planar device in many respects, but some key differences stand out. In particular, the fraction of single-pixel (n1) events is much lower at all energies for the curved devices, and at 5.9 keV there are significant fractions of n5 and n6 events. We attribute this to the difference in thickness: the curved CCDs have twice the depletion depth (100 µm) as the planar one (50 µm), and charge clouds produced by X-rays of all energies experience more charge diffusion and spread to more pixels. This is especially true of soft X-rays that interact near the entrance window, farthest from the collection gates\cite{Milleretal2022c}. The effect is also seen in the C and O K spectra in Figure \ref{fig:94spectra}, where different pixel multiplicities have noticeably different peak energies in the thicker curved devices due to charge lost below threshold in outlying pixels. These variations can be calibrated to reduce the summed FWHM of each curved device to a value very close to that of the planar device at O K, and possibly below it at C K. We conclude that the curving procedure has had no effect on the soft X-ray spectral performance, indicating no significant damage to the backside passivation treatment.

\section{Conclusion}
\label{sect:conclusions}

Future X-ray astrophysics imaging missions will greatly benefit from detectors that closely follow the curved focal surface of high-resolution, grazing-incidence optics. We have demonstrated a method to curve backside-illuminated CCDs to a concave radius of curvature of 2.5 m, similar to that required for the Lynx HDXI, successfully curving three CCDs and showing they meet the required surface conformity of 5 µm RMS. Two devices identified as flight grade from wafer-level testing were subjected to X-ray performance testing, and while there are some interesting but minor differences in performance compared to a planar CCD of the same type, there is no evidence that the curving process has altered the X-ray spectral performance. Noise, gain, dark current, and spectral FWHM are all similar to the planar device and within the range of expected performance for flight devices at notional operating temperatures.

Future work includes subjecting the curved devices to environmental testing, in particular to ensure the mandrel-mounted package is robust to required vibration and shock levels. We will also model the distortion-induced strain to understand the regime within which to expect problems, with an eye toward curving future detectors to tighter tolerance to confirm this modeling.

\acknowledgments

We gratefully acknowledge support from NASA through Astrophysics Research and Analysis (APRA) grant 80NSSC22K0788. We also thank Dr. Kevin A. Grossklaus for assistance with the MBE deposition. 


\end{document}